\documentclass[%
reprint,superscriptaddress,
amsmath,amssymb,
aps,
]{revtex4-2}
\usepackage{hyperref}
\usepackage{graphicx}
\usepackage{dcolumn}
\usepackage{bm}
\usepackage{physics}
\usepackage{amsmath}
\usepackage{color}
\newtheorem{theorem}{Theorem}
\begin{document}
	\title{Improved postselection security analysis of phase error estimation in quantum key distribution}
	\author{Yang-Guang Shan}
	\affiliation{CAS Key Laboratory of Quantum Information, University of Science and Technology of China, Hefei, Anhui 230026, China}
	\affiliation{CAS Center for Excellence in Quantum Information and Quantum Physics, University of Science and Technology of China, Hefei, Anhui 230026, China}
	\author{Zhen-Qiang Yin}
	\email{yinzq@ustc.edu.cn}
	\author{Shuang Wang}
	\email{wshuang@ustc.edu.cn}
	\author{Wei Chen}
	\author{De-Yong He}
	\author{Guang-Can Guo}
	\author{Zheng-Fu Han}
	\affiliation{CAS Key Laboratory of Quantum Information, University of Science and Technology of China, Hefei, Anhui 230026, China}
	\affiliation{CAS Center for Excellence in Quantum Information and Quantum Physics, University of Science and Technology of China, Hefei, Anhui 230026, China}
	\affiliation{Hefei National Laboratory, University of Science and Technology of China, Hefei 230088, China}
	
	\begin{abstract}
	%Quantum key distribution (QKD) enables the generation of secure keys between two distant users. Estimating phase error rate is one of the common method for security analysis of QKD protocols. However, in some protocols, parameter estimation is challenging against general attacks from an eavesdropper. For example, trace distance is a feasible measure for bounding the click rate discrepancy between two quantum states against collective attacks, which cannot be directly applied against coherent attacks. Postselection method can help to extend all security analyses of collective attacks to be against coherent attacks. However, it gives a bad performance. In this article, we propose a method correlating the failure probabilities against collective and coherent attacks, enabling the use of the independent and identically distributed assumption in parameter estimation against coherent attacks. We apply our method to the finite-key analysis of the side-channel-secure (SCS) QKD and the no-phase-postselection (NPP) twin-field (TF) QKD. Our numerical simulations demonstrate that our method can reduce the required number of pulses for SCS QKD by more than an order of magnitude. Additionally, compared to previous analyses based on postselection, our approach significantly improves the performance of NPP TF QKD.
	
	Quantum key distribution (QKD) enables the generation of secure keys between two distant users. Security proof of QKD against general coherent attacks is challenging, while the one against collective attacks is much easier. As an effective and general solution, the postselection method tries to extend security analyses of collective attacks to be against coherent attacks. However, it gives a bad performance. To overcome this drawback, instead of directly calculating key rate by postselection method, we propose a method correlating the failure probabilities of phase error estimation against collective and coherent attacks, enabling the use of the independent and identically distributed assumption in parameter estimation against coherent attacks. Then the key rate can be obtained by uncertainty relation of entropy. Our method can be applied to various QKD protocols, providing better performance compared with the traditional postselection method. For instance, we give the finite-key analyses of the side-channel-secure (SCS) QKD and the no-phase-postselection (NPP) twin-field (TF) QKD to show their performance improvements with the proposed method. 
	
%	However, most QKD protocols require precise preparation of quantum states, necessitating accurate modulation of encoding dimensions of optical pulses while maintaining consistency in non-encoding dimensions across different encodings. Achieving such high precision in practical scenarios is challenging. Luckily, side-channel-secure (SCS) QKD gives a good solution to be immune to the side channels of non-encoding dimensions. SCS QKD ensures security by only requiring bounding the minimal projection probability to the vacuum state. However, the existing security analysis does not give a satisfactory performance under the finite-key case against coherent attacks. In this article, we propose a novel method correlating the failure probabilities of parameter estimations against collective and coherent attacks. By applying this method to the finite-key analysis of SCS QKD, the performance can be significantly improved. In our numerical simulation, the required number of pulses can be reduced by more than one order of magnitude to achieve the same key rate per pulse.
	\end{abstract}
	\maketitle
	\section{introduction}
Quantum key distribution (QKD) \cite{bennett1984quantum} is an art of sharing secure random keys between two distant users (Alice and Bob), using the fundamental principles of quantum mechanics to guarantee security. In recent years, QKD has rapidly advanced both in theoretical developments \cite{lucamarini2018overcoming,ma2018phase,wang2018twin,lin2018simple,cui2019twin,curty2019simple,wang2019practical,zeng2022mode,PRXQuantum.3.020315} and experimental implementations \cite{minder2019experimental,wang2019beating,liu2019experimental,zhong2019proof,fang2020implementation,chen2020sending,zhong2021proof,pittaluga2021600,clivati2022coherent,liu2021field,chen2021twin,wang2022twin,liu2023experimental}.

Over the past few decades, various methods have been proposed to prove the security of different QKD protocols. One of the most commonly used methods is based on the phase error estimation \cite{koashi2009simple,tomamichel2012tight} and the uncertainty relation of entropy \cite{tomamichel2011uncertainty}. This method is easy to apply and performs well, making it a popular choice in many mainstream protocols. 

We review a phase-error security analysis in the following. Firstly we should give the equivalent protocol based on entanglement. In the equivalent protocol, if Alice and Bob measure their local quantum states on a specific basis (referred to as the $\mathbb{Z}$ basis), they will obtain the same key bits as the original protocol. Conversely, if they measure their local quantum states on a complementary basis to the $\mathbb{Z}$ basis (referred to as the $\mathbb{X}$ basis), the error rate, known as the phase error rate, reflects the amount of information leakage to an eavesdropper. Since in most practical protocols, the phase error rate cannot be directly measured, Alice and Bob may send additional states called decoy states \cite{hwang2003quantum,PhysRevLett.94.230504,wang2005beating} to estimate the phase error count. With the known phase error number, we can use the uncertainty relation of entropy \cite{tomamichel2011uncertainty} to bound the min-entropy of Alice conditioned on Eve's knowledge. Finally, with the theorem of quantum leftover hashing \cite{tomamichel2011leftover}, the length of the secure key can be obtained.

Though this kind of analysis seems straightforward, phase error estimation is not always easy. In some simple protocols, the quantum state corresponding to the measurement result of $\mathbb{X}$ basis can be practically prepared or at least can be prepared as a part of a mixed state. In this case, the phase error estimation is relatively easy because every phase error event can be assumed to be independently allocated to be a signal state or a decoy state. However, in some other cases, the $\mathbb{X}$ basis states cannot be prepared. For example, in the side-channel-secure (SCS) protocol \cite{wang2019practical}, Alice and Bob need to estimate the click rate of $\ket{0}_a\ket{\sqrt{\mu}}_b+\ket{\sqrt{\mu}}_a\ket{0}_b$, where $\ket{0}$ represents the vacuum state, $\ket{\sqrt{\mu}}$ is a coherent state of intensity $\mu$, and $a$, $b$ denote the states sent by Alice and Bob respectively. This state cannot be prepared remotely by Alice and Bob. In this case, Alice and Bob can estimate the click rate of a similar state and use this to bound the click rate of the target state. For example, trace distance is an upper bound of the click rate discrepancy between two states \cite{nielsen2010quantum}. However, this method cannot be directly applied against coherent attacks. Constructing inequalities for density matrices might be a viable solution \cite{maeda2019repeaterless}, but it may require preparing additional states. In SCS protocol, Alice and Bob cannot prepare more kinds of states, which will ruin the advantage of side-channel-secure. 

To handle some challenging phase error estimations, some works adopted the postselection method \cite{christandl2009postselection}, which can extend the security analysis against collective attacks to coherent attacks. This approach simplifies phase error estimation by assuming independent and identically distributed attacks for each round. However, the key rate performance from postselection is too conservative, limiting the practicability of the protocol. Moreover, a recent study \cite{nahar2024postselection} pointed out that the original postselection technique cannot be directly applied to prepare-and-measure protocols (including measurement-device-independent (MDI) protocols). This is because most existing security analyses require fixed local ancillas for the sender(s) and these analyses cannot be applied to the case of arbitrarily shared quantum state pairs between the two users, which is required in the original postselection method. Thus, some existing security analyses \cite{liu2020finite,lu2019practical,jiang2024side} based on the postselection method are not rigorous enough.

In this article, we find that a whole postselection method is not necessary for the security analysis. Instead, correlating the phase error estimation against collective and coherent attacks is enough to finish the analysis. We can use the assumption of independent and identical distribution to conduct the parameter estimation against collective attacks. Then we apply the de Finetti reduction with fixed marginal \cite{nahar2024postselection} to extend this parameter estimation to the coherent-attack case. After obtaining the phase error count under coherent attacks, we can finish the security analysis based on the min-entropy calculation \cite{tomamichel2012tight}. 

We apply our method to the SCS QKD. In the numerical simulation, to realize the same key rate per pulse, our method requires more than an order of magnitude fewer pulses compared to previous work. We also apply our method to the no-phase-postselection (NPP) twin-field (TF) QKD \cite{cui2019twin,curty2019simple}, where distinct improvement is also observed.

This article is organized as follows. In Sec. \ref{sec:est} we give our method correlating the parameter estimation against collective and coherent attacks. We apply our method to the SCS protocol in Sec. \ref{sec:scs} and to the NPP TF QKD in \ref{sec:npp}. Finally we conclude our work in Sec. \ref{sec:con}.

\section{parameter estimation with de Finetti reduction \label{sec:est}}
In an analysis against collective attacks, we assume that the eavesdropper Eve applies the same completely positive trace-preserving (CPTP) map to the quantum states of each round. Under this independent scenario, we can assert that if two states are similar and randomly prepared by the sender, they will have similar click rates. This property finds its application in various QKD analyses, for example, in some TF QKD protocols \cite{curty2019simple} and SCS QKD \cite{wang2019practical}. However, it cannot be directly applied in the analyses against coherent attacks. 
%Constructing an inequality of density matrices could be a good solution to this difficulty \cite{maeda2019repeaterless}, but it cannot be applied in protocols like the SCS QKD because it may require preparing more kinds of states in the SCS QKD, which will ruin the advantage of side-channel-secure. 

In this section, we give a method based on the de Finetti reduction with a fixed marginal \cite{nahar2024postselection}. With our method, we can use the same equations (inequalities) to estimate the click rates of any states against coherent attacks, even though these equations (inequalities) come from the analysis against collective attacks. For example, we can still bound the click rate discrepancy between two states with the trace distance even when coherent attacks are conducted. Note that this generalization is not costless, and the failure probability should be increased.

We use a prepare-and-measure protocol to describe our method, but our method can also be applied to MDI-type protocols. In the following, Alice prepares quantum states and sends them to Bob who will measure these states.

In an equivalent protocol based on entanglement, we assume that Alice and Bob conduct the protocol for $N$ rounds. In each round, Alice prepares the state $\rho_{Aa}$, where the subscript $A$ corresponds to the ancillas held by Alice and the subscript $a$ corresponds to the states sent from Alice to Bob. Then the $a$ systems of $\rho_{Aa}^{\otimes N}$ will suffer from a (coherent) CPTP map from the channel and Eve's attack. The states shared by Alice and Bob become $\rho_{A^NB^N}$, where $B$ is the states received by Bob, with the only restriction that $\Tr_B(\rho_{A^NB^N})=\Tr_a(\rho_{Aa})^{\otimes N}$. If we have the assumption of collective attacks, the states shared by Alice and Bob should be $\rho_{AB}^{\otimes N}$ with $\Tr_B(\rho_{AB})=\Tr_a(\rho_{Aa})$.
 
In parameter estimation, Alice and Bob will measure their own part of $\rho_{A^NB^N}$ independently and identically for each round. They should count the numbers of specific measurement results and use these statistics to infer the number of another measurement result. The failure probability of the parameter estimation can be treated as a measurement probability of a specific POVM matrix $M$ operating on $\rho_{A^NB^N}$. 

We give an example to explain the parameter estimation and its failure probability in the following. In a single-photon BB84 protocol, a low bit error rate in the $\mathbb{X}$ basis ($\ket{+}=(\ket{0}+\ket{1})/\sqrt{2}$ and $\ket{-}=(\ket{0}-\ket{1})/\sqrt{2}$) means a low phase error rate in the $\mathbb{Z}$ basis ($\ket{0}$ and $\ket{1}$). We assume that in a hypothetical experiment, Alice and Bob randomly tag each round to be a $\mathbb{X}$ round or a $\mathbb{Z}$ round, but they measure the states of all the rounds on the $\mathbb{X}$ basis. We assume that they find $n_{bit}^X$ errors in $\mathbb{X}$ rounds and $n_{ph}^Z$ errors in $\mathbb{Z}$ rounds. Now Alice will pretend to forget $n_{ph}^Z$ and estimate its upper bound with $n_{bit}^X$. This is because in the real protocol, $\mathbb{Z}$ rounds are measured on the $\mathbb{Z}$ basis and $n_{ph}^Z$ is unknown. The estimated upper bound is denoted as $\bar n_{ph}^Z(n_{bit}^X)$. The failure probability corresponds to the case that $n_{ph}^Z>\bar n_{ph}^Z(n_{bit}^X)$. We can simply define $M$ as a POVM matrix which measures the number of $n_{bit}^X$ and $n_{ph}^Z$ and finds $n_{ph}^Z>\bar n_{ph}^Z(n_{bit}^X)$. Then $\Tr(M\rho_{A^NB^N})$ is the failure probability of the estimation.

%We can simply define $M_x$ as a POVM matrix which finds $n_{bit}^X$ $\mathbb{X}$-basis errors in $\mathbb{X}$ rounds, and define $M_{xz}$ as a projection matrix which finds both $n_{bit}^X$ $\mathbb{X}$-basis errors in the $\mathbb{X}$ rounds and more than $\bar n_{ph}^Z(n_{bit}^X)$ $\mathbb{X}$-basis errors in the $\mathbb{Z}$ rounds. Then a parameter estimation method will give $\Tr(M_{xz}\rho_{A^NB^N})/\Tr(M_x\rho_{A^NB^N})\le \epsilon$ for all $n_{bit}^X$ as the failure probability. However, we do not need $\Tr(M_{xz}\rho_{A^NB^N})/\Tr(M_x\rho_{A^NB^N})\le \epsilon$ for all $n_{bit}^X$ to ensure security. We only need a weaker result that $\sum_{n_{bit}^X}\Tr(M_{xz}\rho_{A^NB^N})\le \epsilon$. Thus the failure probability can be treated as $\Tr(M\rho_{A^NB^N})$, where $M=\sum_{n_{bit}^X}M_{xz}$.

%  we can define $M$ as a projection matrix which finds more than $\bar n_{ph}^Z(n_{bit}^X)$ $\mathbb{X}$-basis errors in the $\mathbb{Z}$ rounds, then $\Tr(M\rho_{A^NB^N})$ is the failure probability of this phase error estimation.

To continue our analysis, we need to define the property of permutation-invariant. For a quantum state composed of $N$ subsystems, we denote $\pi\in S_N$ as a permutation of these $N$ subsystems and $S_n$ includes $N!$ elements. We say an $N$-round state $\rho^N$ is permutation-invariant if $\pi(\rho^N)=\rho^N$ for any $\pi\in S_n$. We also define that a measurement matrix $M$ is permutation-invariant if $\Tr(M\pi(\rho^N))=\Tr(M\rho^N)$ for any $N$-round state $\rho^N$ and any $\pi\in S_n$.

It is easy to find that the measurement matrix of the failure probability $M$ is permutation-invariant, because in the parameter estimation, we only need the numbers of clicks from different states, but we do not care about the positions of these states and clicks.

 We assume that we have found a parameter estimation method against collective attacks, which means $\Tr(M\rho_{AB}^{\otimes N})\le \epsilon$ for all $\rho_{AB}^{\otimes N}$ with $\Tr_B(\rho_{AB})=\Tr(\rho_{Aa})$. Note that almost all existing parameter estimation methods satisfy this requirement, for example, Chernoff bound \cite{chernoff1952measure}, Azuma's inequality \cite{azuma1967weighted} or other concentration inequalities, and so on. The aim is to prove that $\Tr(M\rho_{A^NB^N})\le \epsilon'$ for all $\rho_{A^NB^N}$ with $\Tr_B(\rho_{A^NB^N})=\Tr_a(\rho_{Aa})^{\otimes N}$. Then we can conclude the same parameter estimation result can be used against coherent attacks except an increased failure probability from $\epsilon$ to $\epsilon'$.
 
 We use the de Finetti reduction with a fixed marginal to obtain our result.
 \begin{theorem}
 \cite{nahar2024postselection}Assuming that $\hat\sigma_A$ is a density matrix and $\bar\rho_{A^NB^N}$ is any permutation-invariant extension of $(\hat\sigma_A)^{\otimes N}$. Then there exists a probability measure $\dd \sigma_{AB}$ on the set of non-negative extensions $\sigma_{AB}$ of $\hat\sigma_A$, such that
 \begin{equation}
 	\bar\rho_{A^NB^N}\le g_{N,x}\int \sigma_{AB}^{\otimes N}\dd \sigma_{AB},
 \end{equation}
 where $x=d^2_Ad^2_B$ and $d_A, d_B$ are the dimensions of systems $A, B$ separately. $g_{N,x}=\binom{N+x-1}{N}$.\label{theorem}
 \end{theorem}

In our analysis under coherent attacks, the state $\rho_{A^NB^N}$ is not permutation-invariant. However, we can define that $\bar\rho_{A^NB^N}=\frac{1}{N!}\sum_{\pi\in S_N}\pi(\rho_{A^NB^N})$, which is permutation-invariant. Then we can find that the failure probability of a parameter estimation is the same for $\rho_{A^NB^N}$ and $\bar\rho_{A^NB^N}$ in the following:
\begin{equation}
	\begin{aligned}
	\Tr(M\bar\rho_{A^NB^N})=&\frac{1}{N!}\sum_{\pi\in S_N}\Tr(M\pi(\rho_{A^NB^N}))\\
	=&\frac{1}{N!}\sum_{\pi\in S_N}\Tr(M\rho_{A^NB^N})\\
	=&\Tr(M\rho_{A^NB^N}),
	\end{aligned}
\end{equation}
where the second equality is from the permutation-invariance of $M$. Then using Theorem \ref{theorem}, we have 
\begin{equation}
	\begin{aligned}
	\Tr(M\rho_{A^NB^N})=&\Tr(M\bar\rho_{A^NB^N})\\
	\le& g_{N,x}\int \dd \sigma_{AB} \Tr(M \sigma_{AB}^{\otimes N})\\
	\le& g_{N,x}\epsilon,
	\end{aligned}
\end{equation}
where the first inequality is from Theorem \ref{theorem} and the second inequality is from the assumption that this parameter estimation can be applied against collective attacks with a failure probability $\epsilon$.

Finally, we can conclude that a parameter estimation of the click number of a specific state can be applied against coherent attacks, if it has been proven to be against collective attacks. The failure probability needs to be multiplied by $g_{N,x}$, which can be simplified with $g_{N,x}=\binom{N+x-1}{N}\le(\frac{e(N+x-1)}{x-1})^{x-1}$ \cite{nahar2024postselection}.

Note that this method can also be used in an MDI-type protocol by simply adding the third peer Charlie's system with a dimension of two. 
%In the following sections, we will apply our method to the SCS protocol to show our performance advantage.
 
%The difficulty of the finite-key analysis under coherent attacks comes from the parameter estimation. The phase error of the SCS protocol comes from the clicks of the state $\ket{0}_a\ket{\sqrt{\mu}}_b+\ket{\sqrt{\mu}}_a\ket{0}_b$, where the subscript $a$ corresponds to the state sent by Alice and the subscript $b$ corresponds to the state sent by Bob. In the security analysis under collective attack, the click number of $\ket{0}_a\ket{\sqrt{\mu}}_b+\ket{\sqrt{\mu}}_a\ket{0}_b$ can be estimated by the click numbers of $\ket{0}_a\ket{0}_b$ and $\ket{\sqrt{\mu}}_a\ket{\sqrt{\mu}}_{b}$ from a decomposition
%\begin{equation}
%	\small
%	\ket{0}_a\ket{\sqrt{\mu}}_b+\ket{\sqrt{\mu}}_a\ket{0}_b=c_0\ket{0}_a\ket{0}_b+c_1\ket{\sqrt{\mu}}_a\ket{\sqrt{\mu}}_{b}+c_2\ket{\phi_2}_{ab},\label{eq:decompose}
%\end{equation}
%where $\ket{\phi_2}_{ab}$ is a normalized state. However, for a analysis under coherent attack, a relation of eq.(\ref{eq:decompose}) is not enough to give a relation of the click numbers between these states. In [], the postselection technique is used used  
\section{Application to SCS QKD\label{sec:scs}}
SCS QKD \cite{wang2019practical} is a protocol that can be immune to almost all side channels of the source part. In this protocol, the two users Alice and Bob both act as the source parts. They only need to randomly prepare two types of states, the vacuum state and a weak coherent state. These states can be imperfectly prepared, with the only requirement of the lower bound of the projection probability to the vacuum state. This kind of requirement can be easily met since the upper bounds of pulse intensities are controllable. In the subsequent studies, the problem of imperfect vacuum states is solved \cite{jiang2023side} and a phase-coding SCS protocol is proposed \cite{shan2023practical}. An experimental realization of the SCS protocol has been conducted over a fiber channel of 50 km \cite{zhang2022experimental}, showing its practicability. The existing finite-key analysis is based on the postselection method \cite{jiang2024side}, and we will show the advantage of our method compared with this work.
\subsection{Protocol description of the SCS QKD \label{sec:protocol}}
In this section, we review the process of the SCS protocol and give the specific requirements of the devices to show the property of side-channel-secure.
\begin{enumerate}
	\item 
	\textbf{State preparation.} Alice (Bob) randomly prepares a weak coherent state $\ket{\sqrt{\mu}}$ or a vacuum state $\ket{0}$ with probabilities $p$ and $1-p$ separately. When she (he) chooses to prepare the weak coherent state, she (he) records a classical bit 1 (0) locally, and when she (he) chooses to prepare the vacuum state, she (he) records a classical bit 0 (1) locally. Then they send the states to Charlie who is located in the middle of the channel.
	
	The above description corresponds to the ideal case, but SCS protocol only requires the following to ensure security. For the source parts with side channels, we assume Alice (Bob) prepares a state with a density matrix $\rho_v$ ($\sigma_v$) when she (he) wants to prepare a vacuum state, and she (he) prepares a state with a density matrix $\rho_w$ ($\sigma_w$) when she (he) wants to prepare the weak coherent state. To prove the security of the protocol, we only need the following requirement:
	\begin{equation}
		\begin{aligned}
			\bra{0}\rho_v\ket{0}\ge a_{v0}\ge 0.5,\quad\bra{0}\rho_w\ket{0}\ge a_0\ge 0.5,\\
			\bra{0}\sigma_v\ket{0}\ge b_{v0}\ge 0.5,\quad\bra{0}\sigma_w\ket{0}\ge b_0\ge 0.5,
		\end{aligned}
	\end{equation}
	where $a_{v0}, a_0, b_{v0}, b_0$ are known to Alice and Bob.
	\item 
	\textbf{State measurement.} If Charlie is honest, he will conduct interference measurements on the two pulses from Alice and Bob. He also compensates for the phase shift from the channel to ensure that the two weak coherent states from Alice and Bob will have constructive interference on the left single-photon detector and destructive interference on the right single-photon detector. If only the right detector clicks, Charlie will declare a successful measurement, or he will declare a failed measurement. For simplicity, a successful measurement is also called a click in the following.
	\item 
	\textbf{Post-processing.} After $N$ rounds of the first two steps, we denote an $\mathcal{O}$ event as a click from the case that both Alice and Bob select to prepare the vacuum state, a $\mathcal{B}$ event as a click from the case that both Alice and Bob select to prepare the weak coherent state, and a $\mathcal{Z}$ event as a click from the case that one of Alice and Bob selects to prepare the vacuum state. $n_\mathcal{O}, n_\mathcal{B}, n_\mathcal{Z}$ are the numbers of the corresponding events. Then $n_t=n_\mathcal{O}+n_\mathcal{B}+n_\mathcal{Z}$ is the total click number. 
	
	Alice and Bob conduct error correction to the successful rounds. Then they can know the values of $n_\mathcal{O}$, $n_\mathcal{B}$ and $n_\mathcal{Z}$ because an error only comes from $\mathcal{O}$ and $\mathcal{B}$ events.  With the known $n_\mathcal{O}$ and $n_\mathcal{B}$, Alice and Bob can estimate the upper bound of phase errors. They also know the bit error rate $e_{bit}=(n_\mathcal{B}+n_{\mathcal{O}})/n_t$. Then Alice and Bob conduct privacy amplification to generate the final key.
\end{enumerate}
\subsection{Security analysis of the SCS QKD\label{sec:security}}
In Ref. \cite{jiang2024side}, the authors have proved that the protocol of preparing the imperfect states $\{\rho_v, \rho_w\}$ and $\{\sigma_v, \sigma_w\}$ can be mapped by Eve from a protocol of preparing the perfect states $\{\ket{0},\ket{\sqrt{\mu_A}}\}$ and $\{\ket{0},\ket{\sqrt{\mu_B}}\}$, where
\begin{equation}
	\begin{aligned}
	e^{-\mu_A}=\abs{\sqrt{a_0a_{v0}}-\sqrt{(1-a_0)(1-a_{v0})}}^2,\\
	e^{-\mu_B}=\abs{\sqrt{b_0b_{v0}}-\sqrt{(1-b_0)(1-b_{v0})}}^2.
	\end{aligned}
\end{equation}
Ref. \cite{jiang2024side} also proved that we only need to analyze the security when Alice and Bob prepare the perfect states because this kind of analysis has included the attacks that Eve maps the states $\{\ket{0},\ket{\sqrt{\mu_A}}\}$ ($\{\ket{0},\ket{\sqrt{\mu_B}}\}$) to the states $\{\rho_v, \rho_w\}$ ($\{\sigma_v, \sigma_w\}$) and then attacks it, which equals to attacking the original protocol.

In the equivalent protocol based on entanglement, Alice and Bob prepare the following state:
\begin{equation}
	\small
	\begin{aligned}
\ket{\Phi}=&(1-p)\ket{01}_{AB}\ket{00}_{ab}+p\ket{10}_{AB}\ket{\sqrt{\mu_A}}_a\ket{\sqrt{\mu_B}}_b\\
&+\sqrt{p(1-p)}(\ket{00}_{AB}\ket{0}_a\ket{\sqrt{\mu_B}}_b+\ket{11}_{AB}\ket{\sqrt{\mu_A}}_a\ket{0}_b),
\end{aligned}
\end{equation} 
where the subscripts $A,B$ correspond to the ancillas held by Alice and Bob, and the subscripts $a,b$ correspond to the states sent out by Alice and Bob. If Alice and Bob measure their ancillas on the $\mathbb{Z}$ basis, they can get their local classical bits.

The bits from the $\mathcal{Z}$ events (defined in Section \ref{sec:protocol}) are treated as untagged bits. Thus the key step of the security analysis is to estimate the number of phase errors of these untagged rounds, which are also the bit errors when Alice and Bob measure their ancillas on the $\mathbb{X}$ basis. We define $\ket{++}_{AB}$ and $\ket{--}_{AB}$ as phase errors. Since this definition equals to define $\frac{\ket{++}_{AB}+\ket{--}_{AB}}{\sqrt{2}}=\frac{\ket{00}_{AB}+\ket{11}_{AB}}{\sqrt{2}}$ and $\frac{\ket{++}_{AB}-\ket{--}_{AB}}{\sqrt{2}}=\frac{\ket{01}_{AB}+\ket{10}_{AB}}{\sqrt{2}}$ as phase errors and the latter cannot appear in untagged rounds, in the following we will use the projection probability to $\frac{\ket{00}_{AB}+\ket{11}_{AB}}{\sqrt{2}}$ to calculate the phase error probability.

Under the assumption of collective attacks, Alice and Bob can use the click numbers $n_\mathcal{O}$ and $n_\mathcal{B}$ to estimate the upper bound of the phase error number. With our method shown in section \ref{sec:est}, we can also make the assumption of collective attacks to estimate the phase errors and increase the failure probability to use the result against coherent attacks.

A general collective attack can be treated as a same CPTP map $\mathcal{M}_E$ of Eve for every round, which maps the states sent by Alice and Bob (the $a,b$ states) to a two-dimension state of Charlie indicating a successful measurement or not. For a protocol running for $N$ rounds, the states after Eve's attack is shown as $(\text{id}_{AB}\otimes \mathcal{M}_E\ket{\Phi}\bra{\Phi})^{\otimes N}$.

For any single round of the protocol, the probability that Alice and Bob find a phase error as an untagged round is shown as:
\begin{equation}
	\small
	\begin{aligned}
P_{ph}=&\Tr\bigg(\frac{\ket{00}_{AB}+\ket{11}_{AB}}{\sqrt{2}}\frac{\bra{00}_{AB}+\bra{11}_{AB}}{\sqrt{2}}\otimes\ket{1}_C\bra{1}_C\\
&\text{id}_{AB}\otimes \mathcal{M}_E\ket{\Phi}\bra{\Phi}\bigg)\\
=&\frac{p(1-p)}{2}\Tr\Big(\ket{1}_C\bra{1}_C \mathcal{M}_E\mathcal{P}\left[\ket{0}_a\ket{\sqrt{\mu_B}}_b+\ket{\sqrt{\mu_A}}_a\ket{0}_b\right]\Big),
\end{aligned}\label{eq:ph}
\end{equation}
where $\ket{1}_C$ is Charlie's state indicating a successful measurement, and $\mathcal{P}[\ket{\cdot}]=\ket{\cdot}\bra{\cdot}$. With a same method, we can get the probability of finding an $\mathcal{O}$ event $P_\mathcal{O}$ and the probability of finding a $\mathcal{B}$ event $P_\mathcal{B}$ in the following:
\begin{equation}
	\begin{aligned}
P_{\mathcal{O}}=&\Tr\bigg(\ket{01}_{AB}\bra{01}_{AB}\otimes\ket{1}_C\bra{1}_C\text{id}_{AB}\otimes \mathcal{M}_E\ket{\Phi}\bra{\Phi}\bigg)\\
=&(1-p)^2\Tr\Big(\ket{1}_C\bra{1}_C\mathcal{M}_E(\ket{00}_{ab}\bra{00}_{ab})\Big),
\end{aligned}
\end{equation}
\begin{equation}
	\begin{aligned}
	P_{\mathcal{B}}=&\Tr\bigg(\ket{10}_{AB}\bra{10}_{AB}\otimes \ket{1}_C\bra{1}_C\text{id}_{AB}\otimes \mathcal{M}_E\ket{\Phi}\bra{\Phi}\bigg)\\
	=&p^2\Tr\Big(\ket{1}_C\bra{1}_C\mathcal{M}_E(\ket{\sqrt{\mu_A}}_{a}\ket{\sqrt{\mu_B}}_b\bra{\sqrt{\mu_A}}_{a}\bra{\sqrt{\mu_B}}_b)\Big).
	\end{aligned}
\end{equation}

Then we use the equality of these states from Ref. \cite{jiang2024side}:
\begin{equation}
	\begin{aligned}
	&\ket{0}_a\ket{\sqrt{\mu_B}}_b+\ket{\sqrt{\mu_A}}_a\ket{0}_b\\
	=&c_0\ket{00}_{ab}+c_1\ket{\sqrt{\mu_A}}_a\ket{\sqrt{\mu_B}}_b+\bar{c}_2\ket{\phi_2}_{ab}\\
	=&c_0\ket{\phi_0}+c_1\ket{\phi_1}+\bar c_2\ket{\phi_2},\label{eq:eq}
	\end{aligned}
\end{equation}
where $c_0,c_1>0$ with $c_0c_1=1$ and $\bar{c}_2$ is given to normalize $\ket{\phi_2}_{ab}$ to be 
\begin{equation}
	\bar{c}_2=\sqrt{(c_0+c_1-2e^{-\mu_A/2})(c_0+c_1-2e^{-\mu_B/2})},
\end{equation}
and for simplicity, we use $\ket{\phi_0}, \ket{\phi_1}$ to represent $\ket{00}_{ab}$ and $\ket{\sqrt{\mu_A}}_a\ket{\sqrt{\mu_B}}_b$ separately. Here $c_0$ and $c_1$ can be optimized to realize the best performance, but it is good enough if we take $c_0=e^{-(\mu_A+\mu_B)/4}$ and $c_1=e^{(\mu_A+\mu_B)/4}$ based on experience.

Substitute Eq. (\ref{eq:eq}) into Eq. (\ref{eq:ph}), we get the result in Eq. (\ref{eq:ph0}), where we use Choi's theorem \cite{choi1975completely} to express the CPTP $\mathcal{M}_E$ as $\mathcal{M}_E(\rho)=\sum_iV_{Ei}\rho V_{Ei}^\dagger$ and in the last inequality we use the Cauchy-Schwarz inequality.
\begin{widetext}
\begin{equation}
	\begin{aligned}
	P_{ph}=&\frac{p(1-p)}{2}\Tr\Big(\ket{1}_C\bra{1}_C \mathcal{M}_E\mathcal{P}\left[c_0\ket{\phi_0}+c_1\ket{\phi_1}+\bar{c}_2\ket{\phi_2}\right]\Big)\\
	=&\frac{p(1-p)}{2}\sum_i\Tr\Big(\ket{1}_C\bra{1}_C V_{Ei}\mathcal{P}\left[c_0\ket{\phi_0}+c_1\ket{\phi_1}+\bar{c}_2\ket{\phi_2}\right]V_{Ei}^\dagger\Big)\\
	=&\frac{p(1-p)}{2}\sum_i\abs{\bra{1}_C(c_0V_{Ei}\ket{\phi_0}+c_1V_{Ei}\ket{\phi_1}+\bar{c}_2V_{Ei}\ket{\phi_2})}^2\\
	\le&\frac{p(1-p)}{2}\sum_i\Big(c_0^2\abs{\bra{1}_CV_{Ei}\ket{\phi_0}}^2+c_1^2\abs{\bra{1}_CV_{Ei}\ket{\phi_1}}^2+\bar c_2^2\abs{\bra{1}_CV_{Ei}\ket{\phi_2}}^2\\
	&+2c_0c_1\abs{\bra{1}_CV_{Ei}\ket{\phi_0}}\abs{\bra{1}_CV_{Ei}\ket{\phi_1}}+c_0\bar c_2\abs{\bra{1}_CV_{Ei}\ket{\phi_0}}\abs{\bra{1}_CV_{Ei}\ket{\phi_2}}+c_1\bar c_2\abs{\bra{1}_CV_{Ei}\ket{\phi_1}}\abs{\bra{1}_CV_{Ei}\ket{\phi_2}}\Big)\\
	\le& \frac{p(1-p)}{2}\Bigg(\sum_i(c_0^2\abs{\bra{1}_CV_{Ei}\ket{\phi_0}}^2+c_1^2\abs{\bra{1}_CV_{Ei}\ket{\phi_1}}^2+\bar c_2^2\abs{\bra{1}_CV_{Ei}\ket{\phi_2}}^2)\\
	&\begin{aligned}
	+2c_0c_1\sqrt{\sum_i\abs{\bra{1}_CV_{Ei}\ket{\phi_0}}^2\sum_i\abs{\bra{1}_CV_{Ei}\ket{\phi_1}}^2}&+c_0\bar c_2\sqrt{\sum_i\abs{\bra{1}_CV_{Ei}\ket{\phi_0}}^2\sum_i\abs{\bra{1}_CV_{Ei}\ket{\phi_2}}^2}\\&+c_1\bar c_2\sqrt{\sum_i\abs{\bra{1}_CV_{Ei}\ket{\phi_1}}^2\sum_i\abs{\bra{1}_CV_{Ei}\ket{\phi_2}}^2}\Bigg).
	\end{aligned}
	\end{aligned}\label{eq:ph0}
\end{equation}
\end{widetext}
We can easily find that $P_{\mathcal{O}}=(1-p)^2\sum_i\abs{\bra{1}_CV_{Ei}\ket{\phi_0}}^2$ and $P_{\mathcal{B}}=p^2\sum_i\abs{\bra{1}_CV_{Ei}\ket{\phi_1}}^2$. And for the term of $\ket{\phi_2}$, we can find that $\sum_i\abs{\bra{1}_CV_{Ei}\ket{\phi_2}}^2=\Tr(\ket{1}_C\bra{1}_C\mathcal{M}_E\ket{\phi_2}\bra{\phi_2})\le1$ because both $\ket{1}_C\bra{1}_C$ and $\mathcal{M}_E$ do not increase the trace. Finally, we get a simple upper bound of $P_{ph}$ shown as:
\begin{equation}
\begin{aligned}
	P_{ph}\le& \frac{p(1-p)}{2}\Big(c_0^2\frac{P_{\mathcal{O}}}{(1-p)^2}+c_1^2\frac{P_{\mathcal{B}}}{p^2}+\bar{c}_2^2\\
	&+2c_0c_1\sqrt{\frac{P_\mathcal{O}P_\mathcal{B}}{(1-p)^2p^2}}+c_0\bar{c}_2\sqrt{\frac{P_\mathcal{O}}{(1-p)^2}}+c_1\bar{c}_2\sqrt{\frac{P_\mathcal{B}}{p^2}}\Big).
\end{aligned}
\end{equation}
A same relation has also be given in Refs. \cite{wang2019practical,jiang2023side,jiang2024side}.

Since the phase error estimation is conducted under the assumption of collective attacks, $P_{ph}, P_{\mathcal{O}}$ and $P_{\mathcal{B}}$ are the same for every round. We can use the Chernoff bound of independent variables to estimate the value of $P_{\mathcal{O}}$ and $P_\mathcal{B}$ as $P_\mathcal{O}\le\overline{\text{Cher}}(n_\mathcal{O},\epsilon_0)/N$ and $P_\mathcal{B}\le\overline{\text{Cher}}(n_\mathcal{B},\epsilon_0)/N$, where $\overline{\text{Cher}}(\cdot,\epsilon)$ is the upper bound of the expectation estimated from the observation with a failure probability of $\epsilon$. Then we can use the Chernoff bound to estimate the phase error number as $n_{ph}\le\bar{n}_{ph}= \overline{\text{cher}}(NP_{ph},\epsilon_0)$, where $\overline{\text{cher}}(\cdot,\epsilon)$ is the upper bound of the observation estimated from the expectation with a failure probability $\epsilon$. These upper bounds will be explained in detail in Appendix \ref{app:cher}.

We use the Chernoff bound with a failure probability $\epsilon_0$ three times in our estimation of $\bar n_{ph}$ under collective attacks. With our method given in Section \ref{sec:est}, we can also use a same estimation value of $\bar n_{ph}$ with an increased failure probability $g_{N,x}\times 3\epsilon_0$. Here $x$ is the square of the dimension of Alice, Bob and Charlie. Thus we have $x=d_A^2d_B^2d_C^2=2^2\times 2^2\times 2^2=64$.

With the known phase error number, we can use the method of uncertainty relations of entropy \cite{tomamichel2011uncertainty} to give the final key length against coherent attacks. From the property of the two-universal hash function \cite{tomamichel2011leftover}, the secure key of a length $l$ is $\epsilon_{tot}$ secure, if $\epsilon_{tot}=2\epsilon+\frac{1}{2}\sqrt{2^{l-H_{\min}^\epsilon(Z_t|E')}}$, where $Z_t$ corresponds to Alice's measurement results of her ancillas on the $\mathbb{Z}$ basis of all clicked rounds, and $E'$ is the system of Eve including the information leakage from the error correction step. $H_\text{min}^\epsilon(\cdot|\cdot)$ is the function of conditional smooth min-entropy. Then we can define $l=H_\text{min}^\epsilon(Z_t|E')-2\log_2\frac{1}{2\bar \epsilon}$ with a security parameter of $\epsilon_{tot}=2\epsilon+\bar \epsilon$. 

It is easy to split out the term of information leakage from error correction. We assume that there are $fn_tH_2(e_{bit})$ classical bits published in the error correction, where $f$ is the efficiency of the error correction and $H_2(x)=-x\log_2(x)-(1-x)\log_2(1-x)$ is the binary Shannon entropy. Then a hash of length $\log_2\frac{2}{\epsilon_{cor}}$ is announced for error verification. After passing the error verification, Alice and Bob can assert the identity of their keys with a failure probability $\epsilon_{cor}$. We can redefine $l=H_\text{min}^\epsilon(Z_t|E)-fn_tH_2(e_{bit})-\log_2\frac{2}{\epsilon_{cor}}-2\log_2\frac{1}{2\bar \epsilon}$ with a security parameter $\epsilon_{tot}=2\epsilon+\bar \epsilon+\epsilon_{cor}$ and $E$ is Eve's system before the error correction \cite{tomamichel2012tight}. 

$Z_t$ includes the rounds of $\mathcal{B},\mathcal{O}$ and $\mathcal{Z}$ events, where only the $\mathcal{Z}$ rounds are treated as untagged rounds. We can separate the system $Z_t$ into $Z_{\mathcal{Z}}$ and $Z_{\mathcal{OB}}$ two parts corresponding to the $\mathcal{Z}$ events and $\mathcal{O},\mathcal{B}$ events. With the chain rules of smooth entropy \cite{vitanov2013chain}, we have 
\begin{equation}
	\begin{aligned}
	H_{\text{min}}^\epsilon(Z_t|E)\ge& H_{\text{min}}^{\epsilon_1}(Z_{\mathcal{OB}}|Z_{\mathcal{Z}}E)+H_\text{min}^{\epsilon_2}(Z_\mathcal{Z}|E)-\log_2\frac{2}{\epsilon'^2}\\
	\ge& H_\text{min}^{\epsilon_2}(Z_\mathcal{Z}|E)-\log_2\frac{2}{\epsilon'^2},
	\end{aligned}
\end{equation}
where $\epsilon=\epsilon_2+\epsilon'$ by setting $\epsilon_1=0$.

Using the uncertainty relations for smooth max- and min-entropy \cite{tomamichel2011uncertainty}, the term of smooth min-entropy can be estimated by the smooth max-entropy as follows:
\begin{equation}
	\begin{aligned}
	H_\text{min}^{\epsilon_2}(Z_\mathcal{Z}|E)\ge& n_\mathcal{Z}-H_\text{max}^{\epsilon_2}(X_\mathcal{Z}|B)\\
	\ge& n_\mathcal{Z}-n_\mathcal{Z}H_2(\frac{\bar{n}_{ph}}{n_\mathcal{Z}}),
	\end{aligned}
\end{equation}
where $H_\text{max}^{\epsilon}(\cdot|\cdot)$ is the function of conditional smooth max-entropy, $X_\mathcal{Z}$ corresponds to Alice's ancillas of $\mathcal{Z}$ rounds measured on the $\mathbb{X}$ basis, and the second inequality is given by setting $\epsilon_2=\sqrt{3\epsilon_0g_{N,64}}$ \cite{tomamichel2012tight}.

Finally the key length can be given as:
\begin{equation}
	\small
	l\ge n_\mathcal{Z}-n_\mathcal{Z}H_2(\frac{\bar{n}_{ph}}{n_\mathcal{Z}})-fn_tH_2(e_{bit})-\log_2\frac{2}{\epsilon'^2}-\log_2\frac{2}{\epsilon_{cor}}-2\log_2\frac{1}{2\bar \epsilon},
\end{equation}
with a security parameter $\epsilon_{tot}=\bar\epsilon+\epsilon_{cor}+2\epsilon'+2\sqrt{3\epsilon_0g_{N,64}}$.

\subsection{Numerical simulation of the SCS QKD\label{sec:sim}}
We conduct numerical simulations to show the improvement of our method. To compare with the previous work \cite{jiang2024side}, we use the same parameters shown in table \ref{table:pa}, where $p_d$ is the dark counting rate per pulse of the detectors, $e_d$ is the misalignment error rate, $\eta_d$ is the detecting efficiency of the detectors, $f$ is the efficiency of the error correction, $\alpha_f$ is the fiber loss coefficient (dB/km), and $\epsilon_{tot}$ is the total security parameter.
\begin{table}[h]
	\caption{\label{table:pa}The parameters we used in the simulation.}
	\begin{ruledtabular}
		\begin{tabular}{cccccc}
			$p_d$&$e_d$&$\eta_d$&$f$&$\alpha_f$&$\epsilon_{tot}$\\\hline
			$10^{-9}$&$4\%$&$30.0\%$&$1.1$&$0.2$&$10^{-10}$
		\end{tabular}
	\end{ruledtabular}
\end{table}
\begin{figure}[h]
	\centering
	\includegraphics[width=\columnwidth]{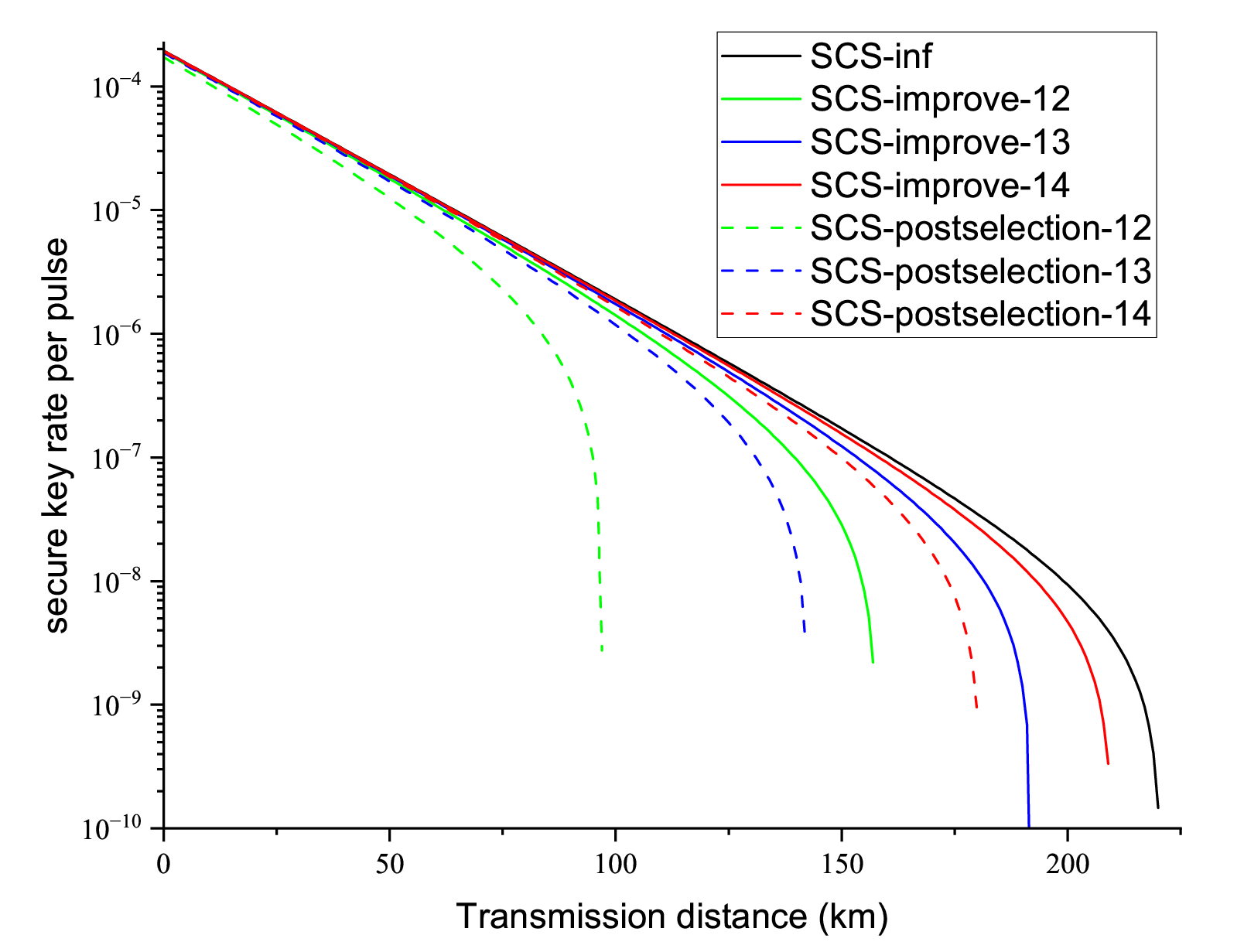}
	\caption{\label{fig:sim}The simulation result of the performance of the SCS protocol. The line SCS-inf corresponds to the asymptotic case with infinite pulses. The line SCS-improve-12 (-13 and -14) is the simulation result with our method based on de Finetti reduction when $10^{12}$ ($10^{13}$ and $10^{14}$) pulses are sent by Alice or Bob. The line SCS-postselection-12 (-13 and -14) is the simulation result from \cite{jiang2024side} with the method of postselection when $10^{12}$ ($10^{13}$ and $10^{14}$) pulses are sent by Alice or Bob.}
\end{figure}

The simulation results are shown in Fig. \ref{fig:sim}, where we simulated the performance of the SCS protocol when the number of pulses sent by Alice (Bob) is $10^{12}, 10^{13}$ and $10^{14}$ separately with our method and the previous analysis. We also simulated the asymptotic case for comparison.

In the simulation result, we can see that the performance of sending $10^{12}$ ($10^{13}$) pulses with our method is even higher than the previous work of sending $10^{13}$ ($10^{14}$) pulses. Thus our method could help a lot to simplify the realization of the SCS protocol by reducing the register requirement and the postprocessing difficulty. The required pulse number to approach the asymptotic case is about $10^{14}$, which is at a practical order of magnitude. 

\section{Application to the NPP TF QKD\label{sec:npp}}
NPP TF QKD \cite{cui2019twin} (also called Curty-Azuma-Lo (CAL) TF QKD \cite{curty2019simple}) is a famous variant of TF QKD. It has a high key rate at a low transmission distance and a similar maximum transmission distance to other TF protocols. In NPP TF QKD, only two phases are used in signal states. Thus these states cannot be treated as mixed states of states with different photon numbers. To estimate the phase error rate, Alice and Bob need to estimate the click number of states $\ket{\sqrt{\mu}}_a\ket{\sqrt{\mu}}_b+\ket{-\sqrt{\mu}}_a\ket{-\sqrt{\mu}}_b$ and $\ket{\sqrt{\mu}}_a\ket{-\sqrt{\mu}}_b+\ket{-\sqrt{\mu}}_a\ket{\sqrt{\mu}}_b$, which cannot be remotely prepared by Alice and Bob with existing technology. Thus in this protocol, Alice and Bob use click rates of phase-randomized coherent states to estimate the phase errors. There are several different methods \cite{lu2019practical,maeda2019repeaterless,curras2021tight} to realize this kind of estimation, including the method based on postselection \cite{lu2019practical}. In this section, we apply our method to the NPP TF QKD and compare its performance with this previous work \cite{lu2019practical} to show the improvement.

\subsection{Protocol description of the NPP TF QKD}
In this section, we review the process of the NPP TF QKD. To simplify the description, our analysis is based on a three-intensity protocol, where the signal state has an intensity $\mu$ and decoy states choose from two intensities $0$ and $\nu$.
\begin{enumerate}
	\item \textbf{State preparation.} Alice (Bob) randomly selects to prepare a signal state with a probably $p$ or a decoy state with a probably $1-p$. If she (he) decides to prepare the signal state, she (he) will randomly select a key bit $s_A (s_B) \in\{0,1\}$ and prepare a coherent state $\ket{\sqrt{\mu}\text{e}^{i\pi s_A}}_a$ ($\ket{\sqrt{\mu}\text{e}^{i\pi s_B}}_b$). If she (he) decides to prepare a decoy state, she (he) will select to prepare a vacuum state $\ket{0}_{a(b)}$ with a probability $p_0$, or prepare a phase-randomized coherent state of an intensity $\nu$ with a probability $1-p_0$. Then Alice and Bob send the states to Charlie, who is located in the middle of the channel. 
	\item \textbf{State measurement.} If Charlie is honest, he will conduct interference measurements on the two pulses from Alice and Bob. He also compensates for the phase shift from the channel to ensure that the two weak coherent states from Alice and Bob will have constructive interference on the left single-photon detector and destructive interference on the right single-photon detector. If only the right detector clicks, Charlie will declare a successful right measurement, and if only the left detector clicks, Charlie will declare a successful left measurement. In other cases, he will declare a failed measurement. For simplicity, a successful right (left) measurement is also called a right (left) click in the following. For the rounds with a right click, Bob flips his corresponding key bits $s_B$.
	
	\item \textbf{Sifting.} After $N$ rounds of the first two steps, Alice and Bob announce their choices of signal states or decoy states of every round. $s_A$ and $s_B$ are kept as sifted bits for the rounds where both Alice and Bob select to send signal states if a click is declared. We denote $n_s$ as the number of sifted bits. For other rounds, Alice and Bob announce their intensity choices.
	
	\item \textbf{Post-processing.} Alice and Bob count the click numbers of different decoy rounds. We denote $n_{00}$ as the number of clicks where both Alice and Bob prepare the vacuum state, $n_{0\nu}$ as the number of clicks where Alice prepares the vacuum state and Bob prepares the coherent state of an intensity $\nu$, and $n_{\nu0}$ as the number of clicks where Alice prepares the coherent state of an intensity $\nu$ and Bob prepares the vacuum state. These numbers are used in the phase error estimation.
	
	Then Alice and Bob conduct error correction and privacy amplification to the sifted bits to get the final key.
\end{enumerate}
\subsection{Security analysis of the NPP TF QKD}
To apply the postselection method or our method to a protocol, the ancillas of Alice and Bob should have finite dimensions. However, when phase randomization is conducted, Alice and Bob need states of infinite dimensions to store infinite phases. We should use the source-map method \cite{nahar2024postselection} to convert the protocol to a finite-dimension version. 

We assume in the real protocol Alice (Bob) prepares the states $\{\rho_i\}$ according to her (his) different choices $i$, and in a virtual protocol she (he) prepares the states $\{\rho_i'\}$ instead. If there exists a CPTP map $M_{\text{CPTP}}$, which maps each of $\rho_i'$ to $\rho_i$, then the security of the virtual protocol implies the security of the real protocol. This lemma is easy to understand since every attack from Eve (denoted as a map $\mathcal{M}_E$) to the real protocol can be applied to the virtual protocol with $M_E\circ M_\text{CPTP}$. In the following, we will give such a virtual protocol of finite dimensions.

In the virtual protocol, if Alice (Bob) chooses to prepare the signal state or the vacuum state, the states are not changed. If Alice (Bob) chooses to prepare the coherent state with an intensity $\nu$, she (he) will prepare the following state,
\begin{equation}
	\bar\rho_{\nu}=\text{e}^{-\nu}\ket{0}\bra{0}+\text{e}^{-\nu}\nu\ket{1}\bra{1}+(1-\text{e}^{-\nu}-\text{e}^{-\nu}\nu)\ket{2_+}\bra{2_+},
\end{equation}
where $\ket{1}$ is the Fock state of single photon, and $\ket{2_+}$ is a state orthogonal to all Fock states. In the real protocol, the prepared state is the phase-randomized coherent state shown as $\rho_\nu=\sum_{j=0}^\infty \text{e}^{-\nu}\nu^j/j!\ket{j}\bra{j}$. We can define that the map $M_\text{CPTP}$ measures $\ket{2_+}\bra{2_+}$ and then prepares $C\sum_{j=2}^\infty \text{e}^{-\nu}\nu^j/j!\ket{j}\bra{j}$ (normalized by $C$). Signal states and the vacuum state cannot be measured to $\ket{2_+}\bra{2_+}$ because $\ket{2+}$ is orthogonal to all Fock states. Thus this map can meet the requirement and we can analyze the security with $\bar\rho_\nu$ in the following.

We give the equivalent protocol based on entanglement in the following. In the equivalent protocol, Alice and Bob prepare the state $\ket{\Phi}=\ket{\Phi}^A\otimes\ket{\Phi}^B$ from Eqs. (\ref{eq:phiA}) and (\ref{eq:phiB}) in the state preparation step, where $\ket{s}_{Ai}, \ket{v}_{Ai}, \ket{\nu}_{Ai}$ correspond to Alice's ancilla storing the choice of a signal state, a vacuum state or $\bar\rho_\nu$. $\ket{0}_{Ap}, \ket{1}_{Ap}, \ket{2}_{Ap}$ correspond to Alice's ancilla storing the photon number of $\bar\rho_\nu$. $\ket{0}_A, \ket{1}_A$ correspond to Alice's ancilla storing the key bit. The states with a subscript $a$ correspond to the states sent out by Alice. The states of Bob are similarly defined.
\begin{widetext}
\begin{equation}
	\begin{aligned}
	\ket{\Phi}^A=&\sqrt{\frac{p}{2}}\ket{s}_{Ai}\ket{0}_{Ap}\left(\ket{0}_A\ket{\sqrt{\mu}}_a+\ket{1}_A\ket{-\sqrt{\mu}}_a\right)
	+\sqrt{(1-p)p_0}\ket{v}_{Ai}\ket{0}_{Ap}\ket{0}_A\ket{0}_a\\
&+\sqrt{(1-p)(1-p_0)}\ket{\nu}_{Ai}\ket{0}_A
	\left(\sqrt{\text{e}^{-\nu}}\ket{0}_{Ap}\ket{0}_a+\sqrt{\text{e}^{-\nu}\nu}\ket{1}_{Ap}\ket{1}_a
	+\sqrt{1-\text{e}^{-\nu}-\text{e}^{-\nu}\nu}\ket{2}_{Ap}\ket{2_+}_a\right),
	\end{aligned}\label{eq:phiA}
\end{equation}
\begin{equation}
	\begin{aligned}
		\ket{\Phi}^B=&\sqrt{\frac{p}{2}}\ket{s}_{Bi}\ket{0}_{Bp}\left(\ket{0}_B\ket{\sqrt{\mu}}_b+\ket{1}_B\ket{-\sqrt{\mu}}_b\right)
		+\sqrt{(1-p)p_0}\ket{v}_{Bi}\ket{0}_{Bp}\ket{0}_B\ket{0}_b\\
		&+\sqrt{(1-p)(1-p_0)}\ket{\nu}_{Bi}\ket{0}_B
		\left(\sqrt{\text{e}^{-\nu}}\ket{0}_{Bp}\ket{0}_b+\sqrt{\text{e}^{-\nu}\nu}\ket{1}_{Bp}\ket{1}_b
		+\sqrt{1-\text{e}^{-\nu}-\text{e}^{-\nu}\nu}\ket{2}_{Bp}\ket{2_+}_b\right).\label{eq:phiB}
	\end{aligned}
\end{equation}
\end{widetext}

The dimension of Alice's (Bob's) ancillas is six. Thus considering Charlie's system of dimension three (indicating a left click, a right click, or no click), the total dimension of this protocol is $108$.

In the equivalent protocol, a phase error corresponds to the clicks with a measurement result of $\ket{ss}_{AiBi}\ket{00}_{ApBp}\ket{++}_{AB}$ and $\ket{ss}_{AiBi}\ket{00}_{ApBp}\ket{--}_{AB}$. However, evaluating phase-correct events is more convenient, which corresponds to $\ket{ss}_{AiBi}\ket{00}_{ApBp}\ket{+-}_{AB}$ and $\ket{ss}_{AiBi}\ket{00}_{ApBp}\ket{-+}_{AB}$. In a phase-correct estimation against collective attacks, the probability of a phase-correct event for each round is shown in Eq. (\ref{eq:phnpp}).
\begin{widetext}
\begin{equation}
	\begin{aligned}
	P_{cor}=&\frac{p^2}{4}\Tr((\ket{+-}\bra{+-}_{AB}+\ket{-+}\bra{-+}_{AB})\otimes\ket{1}\bra{1}_C \text{id}_{AB}\otimes\mathcal{M}_E\mathcal{P}((\ket{0}_A\ket{\sqrt{\mu}}_a+\ket{1}_A\ket{-\sqrt{\mu}}_a)(\ket{0}_B\ket{\sqrt{\mu}}_b+\ket{1}_B\ket{-\sqrt{\mu}}_b)))\\
	=&\frac{p^2}{4}\Tr(\text{id}_{AB}\otimes\mathcal{M}_E(\mathcal{P}((\frac{\ket{\sqrt\mu,\sqrt\mu}_{a,b}-\ket{-\sqrt\mu,-\sqrt\mu}_{a,b}}{\sqrt{2}})+\mathcal{P}(\frac{\ket{\sqrt\mu,-\sqrt\mu}_{a,b}-\ket{-\sqrt\mu,\sqrt\mu}_{a,b}}{\sqrt{2}}))))\\
	=&p^2\Tr(\text{id}_{AB}\otimes\mathcal{M}_E(\mathcal{P}(\sum_{j,k=0}^\infty\sqrt{\text{e}^{-2\mu}\frac{\mu^{2j+2k+1}}{(2j)!(2k+1)!}}\ket{2j}_a\ket{2k+1}_b)+\mathcal{P}(\sum_{j,k=0}^\infty\sqrt{\text{e}^{-2\mu}\frac{\mu^{2j+2k+1}}{(2j+1)!(2k)!}}\ket{2j+1}_a\ket{2k}_b)))
	\end{aligned}\label{eq:phnpp}
\end{equation}

In our three-intensity protocol, we only care about the terms of $\ket{01}_{ab}$ and $\ket{10}_{ab}$, so we can define $\sum_{j,k=0}^\infty\sqrt{\text{e}^{-2\mu}\frac{\mu^{2j+2k+1}}{(2j)!(2k+1)!}}\ket{2j}_a\ket{2k+1}_b=\sqrt{\text{e}^{-2\mu}\mu}\ket{01}_{ab}+\sqrt{\text{e}^{-2\mu}(\sinh(\mu)\cosh(\mu)-\mu)}\ket{e-o}_{ab}$ and $\sum_{j,k=0}^\infty\sqrt{\text{e}^{-2\mu}\frac{\mu^{2j+2k+1}}{(2j+1)!(2k)!}}\ket{2j+1}_a\ket{2k}_b=\sqrt{\text{e}^{-2\mu}\mu}\ket{10}_{ab}+\sqrt{\text{e}^{-2\mu}(\sinh(\mu)\cosh(\mu)-\mu)}\ket{o-e}_{ab}$, where $\ket{o-e}$ and $\ket{e-o}$ are normalized. Then the phase-correct probability can be bounded with the same method of Eq. (\ref{eq:ph0}) shown below.

	\begin{equation}
		\begin{aligned}
			P_{cor}=&p^2\Tr\Big(\text{id}_{AB}\otimes\mathcal{M}_E(\mathcal{P}(\sqrt{\text{e}^{-2\mu}\mu}\ket{01}_{ab}+\sqrt{\text{e}^{-2\mu}(\sinh(\mu)\cosh(\mu)-\mu)}\ket{e-o}_{ab})\\
			&+\mathcal{P}(\sqrt{\text{e}^{-2\mu}\mu}\ket{10}_{ab}+\sqrt{\text{e}^{-2\mu}(\sinh(\mu)\cosh(\mu)-\mu)}\ket{o-e}_{ab}))\Big)\\
			\ge& p^2 \left(\sqrt{\text{e}^{-2\mu}\mu}\sqrt{\Tr(\text{id}_{AB}\otimes\mathcal{M}_E\ket{01}\bra{01}_{ab})}-\sqrt{\text{e}^{-2\mu}(\sinh(\mu)\cosh(\mu)-\mu)}\sqrt{\Tr(\text{id}_{AB}\otimes\mathcal{M}_E\ket{e-o}\bra{e-o}_{ab})}\right)^2\\
			&+p^2 \left(\sqrt{\text{e}^{-2\mu}\mu}\sqrt{\Tr(\text{id}_{AB}\otimes\mathcal{M}_E\ket{10}\bra{10}_{ab})}-\sqrt{\text{e}^{-2\mu}(\sinh(\mu)\cosh(\mu)-\mu)}\sqrt{\Tr(\text{id}_{AB}\otimes\mathcal{M}_E\ket{o-e}\bra{o-e}_{ab})}\right)^2
		\end{aligned}\label{eq:cor}
	\end{equation}

Here $\Tr(\text{id}_{AB}\otimes\mathcal{M}_E\ket{01}\bra{01}_{ab})$ corresponds to the click rate of $\ket{01}_{ab}$ and $\Tr(\text{id}_{AB}\otimes\mathcal{M}_E\ket{10}\bra{10}_{ab})$ corresponds to the click rate of $\ket{10}_{ab}$. If the click rate of $\ket{01}_{ab}$ is known and $\sqrt{\text{e}^{-2\mu}\mu}\sqrt{\Tr(\text{id}_{AB}\otimes\mathcal{M}_E\ket{01}\bra{01}_{ab})}>\sqrt{\text{e}^{-2\mu}(\sinh(\mu)\cosh(\mu)-\mu)}$, the lower bound of the first term is given when $\Tr(\text{id}_{AB}\otimes\mathcal{M}_E\ket{e-o}\bra{e-o}_{ab})=1$, or the lower bound is $0$. The same result holds for the second term.
\end{widetext}

Then we should use the decoy states to estimate the click rates of the states $\ket{01}_{ab}$ and $\ket{10}_{ab}$. Since our parameter estimation is under the assumption of collective attacks, we can easily get the click rate lower bounds with Chernoff bound in the following. Here for simplicity we define $\Tr(\text{id}_{AB}\otimes\mathcal{M}_E\ket{mn}\bra{mn}_{ab})=q_{mn}$ and  $\Tr(\text{id}_{AB}\otimes\mathcal{M}_E\ket{0}_a\ket{2_+}_b\bra{0}_a\bra{2_+}_{b})=q_{02_+}$. $q_{2_+0}$ is similarly defined.
\begin{widetext}
\begin{equation}
	N(1-p)^2p_0^2q_{00}\le \overline{\text{Cher}}(n_{00},\epsilon_0),
\end{equation}
\begin{equation}
	\begin{aligned}
		\text{e}^{-\nu}q_{00}+\text{e}^{-\nu}\nu q_{01}+(1-\text{e}^{-\nu}-\text{e}^{-\nu}\nu)q_{02_+}\ge \frac{\underline{\text{Cher}}(n_{0\nu},\epsilon_0)}{N(1-p)^2p_0(1-p_0)},\\
		\text{e}^{-\nu}q_{00}+\text{e}^{-\nu}\nu q_{10}+(1-\text{e}^{-\nu}-\text{e}^{-\nu}\nu)q_{2_+0}\ge \frac{\underline{\text{Cher}}(n_{\nu0},\epsilon_0)}{N(1-p)^2p_0(1-p_0)},
		\end{aligned}
\end{equation}
\begin{equation}
	\begin{aligned}
	q_{01}\ge\frac{1}{\text{e}^{-\nu}\nu}(\frac{\underline{\text{Cher}}(n_{0\nu},\epsilon_0)}{N(1-p)^2p_0(1-p_0)}-\text{e}^{-\nu}\frac{\overline{\text{Cher}}(n_{00},\epsilon_0)}{N(1-p)^2p_0^2}-(1-\text{e}^{-\nu}-\text{e}^{-\nu}\nu)),\\
	q_{10}\ge\frac{1}{\text{e}^{-\nu}\nu}(\frac{\underline{\text{Cher}}(n_{\nu0},\epsilon_0)}{N(1-p)^2p_0(1-p_0)}-\text{e}^{-\nu}\frac{\overline{\text{Cher}}(n_{00},\epsilon_0)}{N(1-p)^2p_0^2}-(1-\text{e}^{-\nu}-\text{e}^{-\nu}\nu)).
	\end{aligned}
\end{equation}
\end{widetext}

With the known $P_{cor}$, we can estimate the lower bound of phase correct events against collective attacks as $n_{cor}\ge\underline{\text{cher}}(NP_{cor},\epsilon_0)$. In the estimation, the total failure probably is $4\epsilon_0$. With our method shown in Sec. \ref{sec:est}, the same estimation holds against coherent attacks with a failure probably $g_{N,108^2}\times 4\epsilon_0$.

With the same method shown in Sec. \ref{sec:security}, the final key length can be given as:
\begin{equation}
	l\ge n_s(1-H_2(1-\frac{\underline{n}_{cor}}{n_s})-fH_2(e_{bit}))-\log_2\frac{2}{\epsilon_{cor}}-2\log_2\frac{1}{2\bar \epsilon},
\end{equation}
with a security parameter $\epsilon_{tot}=\bar \epsilon+\epsilon_{cor}+2\sqrt{4\epsilon_0g_{N,108^2}}$.
\subsection{Numerical simulation of the NPP TF QKD}
We conduct numerical simulation to compare the performance from our method and the original postselection method \cite{lu2019practical}. The parameters we used in the simulation are the same as shown in Table \ref{table:pa}. Note that in Ref. \cite{lu2019practical}, the dimension of the protocol is wrong and we fixed it to be $108$ in the simulation. For the fairness of the comparison, we also use a three-intensity scheme in the simulation based on postselection.

\begin{figure}[h]
	\centering
	\includegraphics[width=\columnwidth]{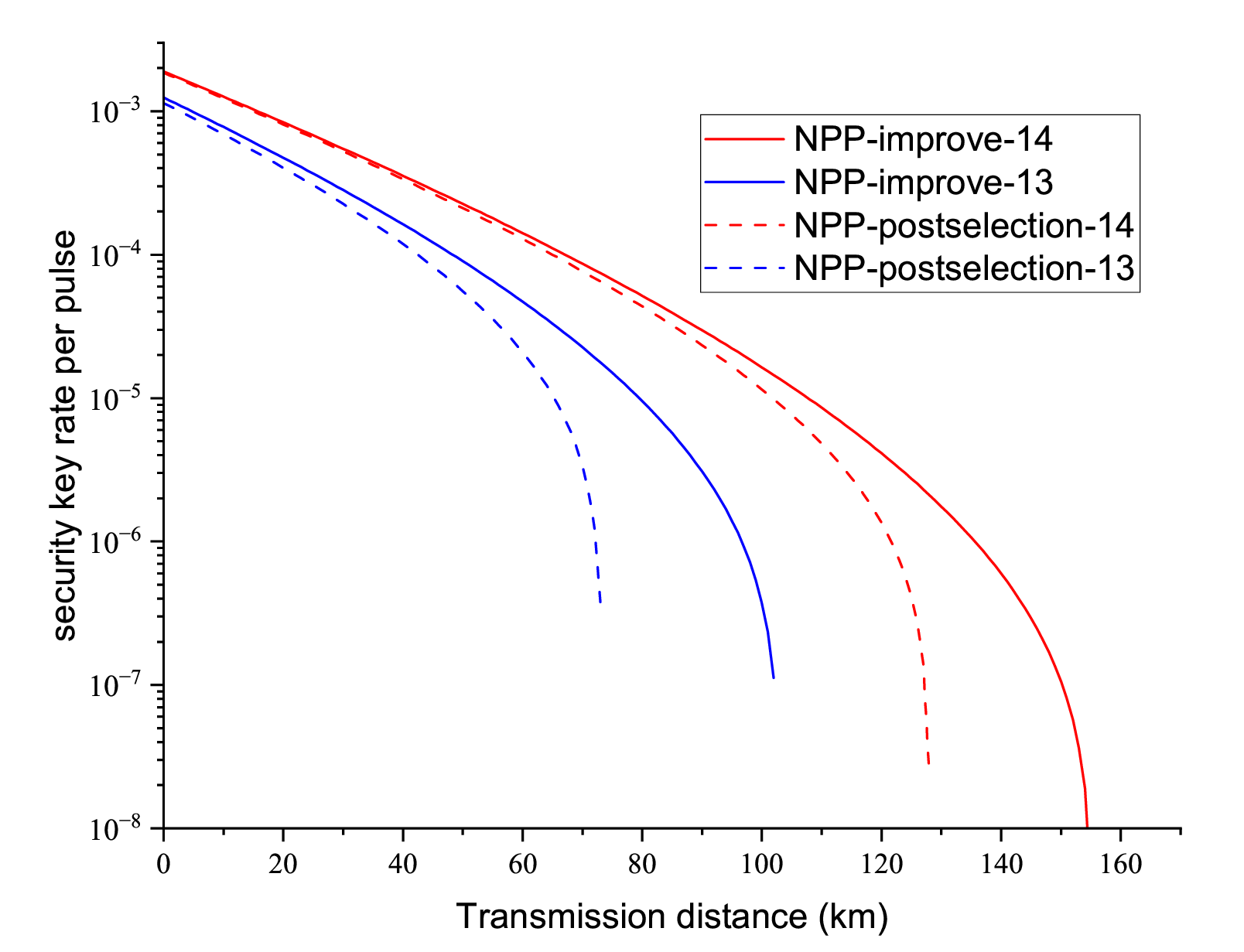}
	\caption{\label{fig:simnpp}The simulation result of the performance of the NPP TF QKD. The line NPP-improve-13 (-14) is the simulation result with our method based on de Finetti reduction when $10^{13}$ ($10^{14}$) pulses are sent by Alice or Bob. The line NPP-postselection-13 (-14) is the simulation result with the original postselection method when $10^{13}$ ($10^{14}$) pulses are sent by Alice or Bob.}
\end{figure}

The simulation result is shown in Fig. \ref{fig:simnpp}. Our simulation shows that our method can drastically improve the performance of the NPP TF QKD compared with the original postselection method. Note that we only show the performance improvement over the postselection method here, and other security analyses based on different methods can realize better performances \cite{maeda2019repeaterless,curras2021tight}.

\section{conclusion\label{sec:con}}
We propose a new method to connect the parameter estimation against collective and coherent attacks. Thus in the security proof against coherent attacks, we can use the assumption of independent identical distribution.

We apply our method to the security analysis of the SCS QKD. In the numerical simulation, we find that the key rate per pulse based on our method is higher than the previous work, even when only one-tenth pulses are sent in our method. As far as we know, our method has the best performance among existing works for SCS QKD.

We apply our method to the security analysis of the NPP TF QKD. In the numerical simulation, our analysis can help to improve the performance compared with the previous work based on postselection. 

Our method may be helpful to other protocols that are hard to analyze against coherent attacks, for example, the finite-key analysis for the discrete-phase-randomization protocols \cite{wang2023finite}. Note that our method can be further improved using the de Finetti reduction with other symmetries \cite{nahar2024postselection}, for example, the block-diagonal symmetry, which may help to decrease the coefficient $g_{N,x}$.
\begin{acknowledgments}
	This work has been supported by the National Key Research and Development Program of China (Grant No. 2020YFA0309802), the National Natural Science Foundation of China (Grant Nos. 62171424, 62271463 and 62301524), Prospect and Key Core Technology Projects of Jiangsu provincial key R \& D Program (BE2022071), the Fundamental Research Funds for the Central Universities, the Innovation Program for Quantum Science and Technology (Grant No. 2021ZD0300701), the Natural Science Foundation of Anhui (No.2308085QF216), the China Postdoctoral Science Foundation (2022M723064).
\end{acknowledgments}
\appendix
\section{Chernoff bound\label{app:cher}}
The Chernoff bound \cite{chernoff1952measure,mitzenmacher2017probability} is widely used in the analysis of QKD protocol. In the following, we give the content of it.

\textbf{Multiplicative Chernoff bound.} Suppose $X_1, X_2,\dots,X_n$ are independent Bernoulli random variables and let $X=\sum_{i=1}^nX_i$. $E$ is the expectation value of $X$. Then we have 
\begin{equation}
	\text{Pr}(X\ge(1+\xi_1)E)\le e^{-\xi_1^2E/(2+\xi_1)}
\end{equation}
for $\xi_1\ge0$, and 
\begin{equation}
	\text{Pr}(X\le(1-\xi_2)E)\le e^{-\xi_2^2E/2}
\end{equation}
for $0<\xi_2<1$.

By solving $e^{-\xi_1^2E/(2+\xi_1)}=e^{-\xi_2^2E/2}=\epsilon_x$, we can get the upper and lower bounds of $X$ shown as 
\begin{equation}
	\begin{aligned}
		X\le\overline{\text{cher}}(E,\epsilon_x)=&E+\frac{1}{2}\ln\frac{1}{\epsilon_x}+\frac{1}{2}\sqrt{\ln^2\frac{1}{\epsilon_x}+8E\ln\frac{1}{\epsilon_x}},	\\
		X\ge\underline{\text{cher}}(E,\epsilon_x)=&E-\sqrt{2E\ln\frac{1}{\epsilon_x}},
	\end{aligned}\label{cher}
\end{equation}
with failure probability $\epsilon_x$ separately.

When $X$ is known but $E$ is unknown, we can also estimate the bound of $E$ by solving $E$ from eq.(\ref{cher}) in the following with a failure probability $\epsilon_x$ separately.
\begin{equation}
	\begin{aligned}
		E\le\overline{\text{Cher}}(X,\epsilon_x)=X+\ln\frac{1}{\epsilon_x}+\sqrt{\ln^2\frac{1}{\epsilon_x}+2X\ln\frac{1}{\epsilon_x}}\\
		E\ge\underline{\text{Cher}}(X,\epsilon_x)=X+\frac{1}{2}\ln\frac{1}{\epsilon_x}-\frac{1}{2}\sqrt{\ln^2\frac{1}{\epsilon_x}+8X\ln\frac{1}{\epsilon_x}}
	\end{aligned}
\end{equation}

	\bibliography{refe.bib}
\end{document}